\documentclass[aps,prf,10pt,superscriptaddress,showpacs,showkeys]{revtex4}

\usepackage{graphicx}

\usepackage{hyperref}
\usepackage{setspace}
\usepackage{amsmath}
\usepackage{amssymb}
\usepackage{xcolor}
\usepackage{soul}
\usepackage{siunitx}
\usepackage[normalem]{ulem}

\renewcommand{\paragraph}[1]{\emph{#1} --}
\usepackage{float}

\begin{document}

\title{Rapid Viscoelastic Spreading}

\author{Ambre Bouillant}
\author{Pim J. Dekker}
\author{Michiel A. Hack}
\author{Jacco H. Snoeijer}

\affiliation{Physics of Fluids Group, Mesa+ Institute, University of Twente, 7500 AE Enschede, The Netherlands}

\date{\today}

\begin{abstract}
We investigate the rapid spreading dynamics of a viscoelastic drop on a solid. Upon contact, surface tension drives a fast motion of the contact line along the substrate. Here, we resolve this motion for viscoelastic liquids by experiments on aqueous polymer solutions of varying concentration. It is found that while the contact line motion is only mildly affected by the polymers, the interface profile becomes much sharper owing to singular polymer stress. This behaviour is reminiscent of that observed for viscoelastic drop coalescence; in this light we revisit the analogy between spreading and coalescence, which sheds new light also on the spreading of Newtonian liquids. 
\end{abstract}

\maketitle

\section{Introduction}

When a drop comes into contact with a surface, it starts to spread on the solid.  A liquid bridge connecting the drop to the substrate moves radially outwards from the contact point, wetting a circular area of radius $r_0(t)$.  The fate of the drop after contact has been the subject of numerous studies, focussing both on static and dynamical aspects. The statics deals with the final shape of the drop and ultimate spreading extent, while the dynamics is dedicated to how a contact line moves and how a drop reaches its final state \cite{PGG1985wetting,Bonn2009wetting,Snoeijer2013moving}.
In partial wetting conditions, the equilibrium contact angle $\theta$ sets the final shape of the liquid drop. For the most common case of water drops, the combination of limited spreading and low viscosity of the liquid means that the spreading dynamics is resisted by inertia. A balance between the driving capillary effect with the opposing inertia provides the relevant time scale, namely, the inertio-capillary time $\tau=\sqrt{\rho R^3/\gamma}$,  where $\gamma$ denotes the liquid surface tension and $\rho$, the liquid density. For a water drop with millimetric size $R$, $\tau \approx 3$~ms. When observed with the naked eye, the drop thus reaches it final shape quasi-instantaneously. In complete wetting conditions, by contrast, the liquid keeps on spreading slowly making it easily observable. This late stage in the spreading of Newtonian liquids was first studied by Tanner \cite{Tanner1979}, who showed that the radius of the wetted area $r_0(t)$ grows as $t^{1/10}$. This slow regime is dominated by strong viscous dissipation close to the moving contact line \cite{Tanner1979,Voinov1976,Cox1986}. This law, however, fails to capture the early stages of the spreading dynamics, where the geometry of the liquid bridge is manifestly different from the slender wedge assumption that underlies Tanner's law. 

The rapid spreading that occurs immediately after the drop is brought into contact with spreading was addressed much more recently, starting with Biance \textit{et al.}~\cite{Biance2004}. They identified a regime at early time ($t<\tau$), where inertia balances capillarity. The size of the wetted area $r_0(t)$ grows as 
\begin{equation}\label{eq:scalinglaw}
\frac{r_0}{R} \sim \left( \frac{\gamma}{\rho R^3} \right)^{1/4}\, t^{1/2}  = \left( \frac{t}{\tau}\right)^{1/2},
\end{equation}
which involves the inertio-capillary time. This initial work focussed on hydrophilic substrates that are completely wetting, and was followed by other studies that varied the substrate properties using both experiments and simulations \cite{Bird2008,Carlson2012,Carlson2012EPL,Winkels2012,Eddi2013b,Stapelbroek2014}. 
While some works indicated that properties of the substrate influence the dynamics \cite{Bird2008,Carlson2012,Carlson2012EPL}, it was found that -- at very early times -- the spreading dynamics is universal \cite{Winkels2012,Eddi2013b} and described by the same scaling (\ref{eq:scalinglaw}) as for hydrophilic surfaces. Surprisingly, this universal spreading was found to be irrespective of the substrate wettability and nature (whether it be completely or partially wetting, or even soft), and despite the presence of chemical patterns or any roughness (anisotropic, random or even controlled) \cite{Stapelbroek2014}. 

Curiously enough, the spreading law (\ref{eq:scalinglaw}) is identical to the growth of the bridge of during the coalescence of two drops, in the inviscid regime \cite{Eggers1999, Duchemin2003, thoroddsen2005, Aarts2005,Paulsen2011,Sprittles2014}. The analogy between rapid spreading and coalescence has proven very robust, both in the low-viscosity limit \cite{Biance2004,Courbin2009,Winkels2012} and in the viscous limit \cite{Eddi2013b}. Indeed, when a drop is brought into contact with a solid wall or with another drop, in both cases the dynamics evolves through a singular liquid bridge that is vanishingly small at the moment of contact. The success of the spreading-coalescence analogy has been attributed to the fact that the coalescence geometry exhibits a mirror-plane that mimics the role of the substrate in spreading. One should bear in mind, however, that it is not obvious that the flow along the plane of symmetry  is similar in both cases, as for the spreading over a substrate one expects the no-slip condition to impede the motion of the contact line \cite{Carlson2012,Carlson2012EPL,Johansson2018}. Still, the spreading-coalescence analogy offers a rationalisation as to why rapid drop spreading of Newtonian liquids is insensitive to substrate properties.

In the present study we investigate the rapid spreading of viscoelastic liquids, focusing on polymer suspensions. Viscoelastic liquids are hybrid in nature: they can flow like ordinary liquids, yet respond elastically when excited by rapid deformations \cite{tanner2000engineering}.
This is illustrated by saliva and most biological fluids that contain long and flexible polymers that are viscoelastic in nature. When a drop of saliva sandwiched between two fingers is drawn away, it can stretch by several times its radius without breaking, an elastic response that leads to a highly elongated threads \cite{anna2001}. However, when loaded over a long timescale, such a drop flows as a liquid and spreads out on the substrate. This viscoelastic behaviour originates from the polymer chains that are suspended in the Newtonian solvent. At rest, the polymer chains tend to form coils, for which the free energy is minimal. When stretched by the flow, however, the polymers elongate and exert an additional stress onto the solvent. Once the stretching is turned off, polymers relax back to the coil-like structure, which comes with a charactersistic relaxation time $\lambda$. Viscoelastic liquids are ubiquitous in biofluids and in technologies such as coating, printing, aerosol generation and polymer processing, and their flow raises many challenges. Of particular interest is how viscoelastic liquids behave near singularities \cite{renardy2000mathematical}, such as flows around sharp edges \cite{hinch1993, renardy1993, evans2008}, bubble cusps \cite{astarita1965, joseph1995, jimmy_annual2018} or during the breakup of drops \cite{ bazilevskii1990, entov1997jnnfm, anna2001, amarouchene2001prl, clasen2006,eggers-2020-jfm,bonn2020}. Such flows involve regions of extreme polymer stretching -- as is also expected to be the case for rapid viscoelastic spreading. 

\begin{figure*}[t]
	\centering\includegraphics[width=0.8\textwidth]{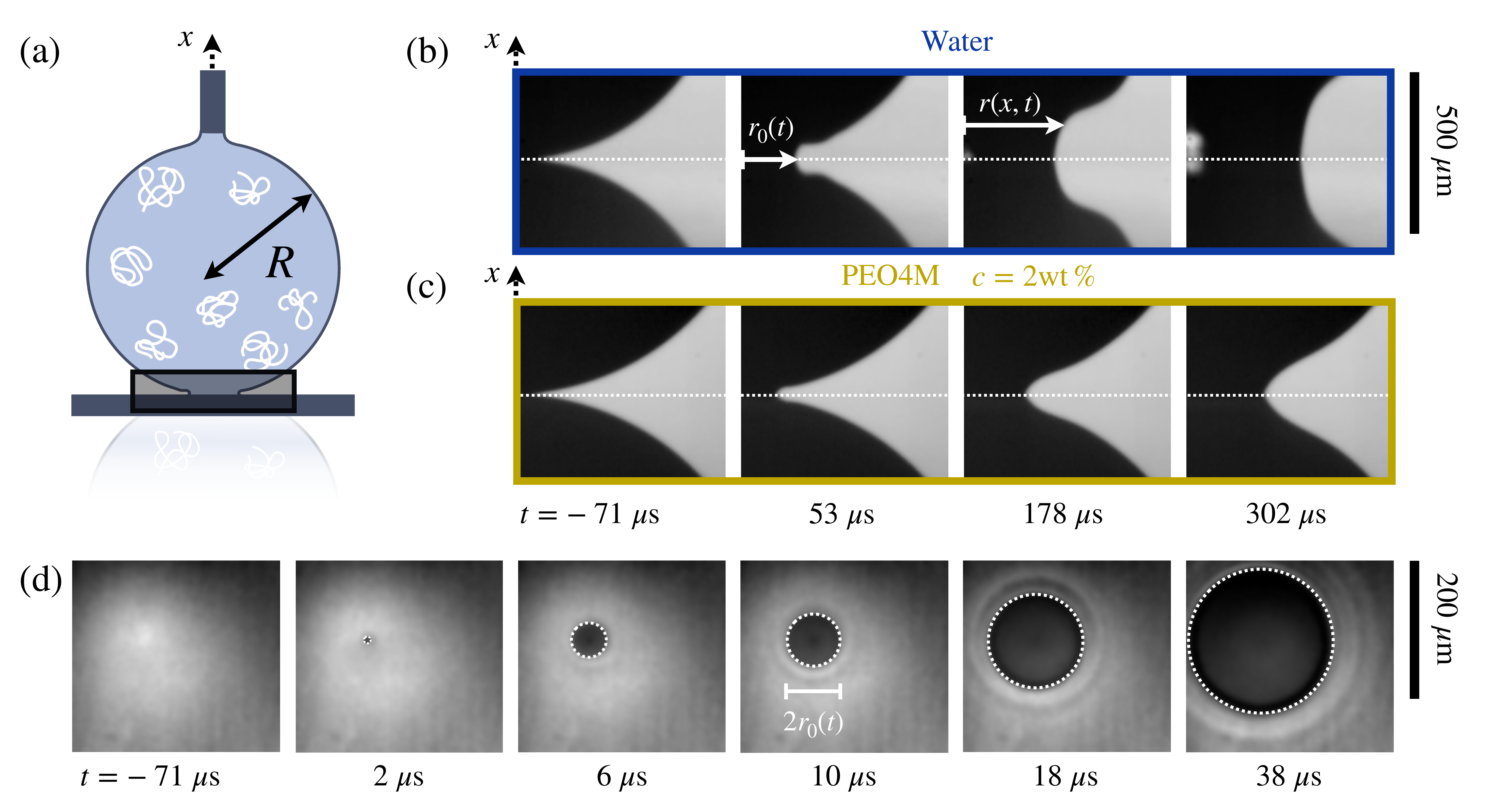}%
	\caption{(a) Side-view schematic of the experimental setup: a drop of PEO is slowly brought into contact with a smooth, transparent and clean glass plate. (b) Upon contact, a water drop rapidly spreads over the solid. The white dashed line indicates the position of the substrate; the lower half of the image is a reflection. The contact line motion is described by the rapid growth of the contact radius $r_0$. (c) Spreading of a 2.0 wt\% PEO drop. The wetting dynamics is weakly slowed down and the interface profiles are much sharper, as compared to the case of water. The scale bar, common to both series of snapshots, indicates 500~$\mu$m. (d) Series of snapshots taken from below of a drop spreading on a (transparent) substrate. Prior to contact, the thin air gap separating the drop from the solid produces an interference pattern. After contact, the liquid quickly wets the solid: the dark area expands radially by typically 200~$\mu$m in 50~$\mu$s, resulting in spreading velocities of about 5~m/s. The scale bar indicate 200~$\mu$m.}
	\label{fig1}
\end{figure*}

Here we reveal how the addition of polymers influences the early stages of drop  spreading on a solid substrate. Given the spreading-coalescence analogy discussed above, the coalescence of viscoelastic drops naturally serves as a benchmark for rapid viscoelastic spreading \cite{varma2020,Dekker2021}. Surprisingly, the bridge temporal evolution during coalescence was found to be only mildly affected by the addition of polymer \cite{Dekker2021}: the bridge growth is nearly identical to that for  Newtonian fluids of low viscosity \cite{Eggers1999, Duchemin2003,thoroddsen2005, Aarts2005,Ristenpart2006, Paulsen2011, Eddi2013}, and was observed to follow (\ref{eq:scalinglaw}) for spherical drops. However, the spatial features of the coalescing bridges are modified: profiles are sharper and the bridge curvature is enhanced. In the present study we experimentally probe the rapid spreading of polymeric drops. Specifically, we resolve the spatio-temporal dynamics of the bridge growth, and inquire whether the spreading-coalescence analogy is robust in the context of viscoelastic liquids.

\section{Experimental procedure}

Spreading experiments are performed on partially wetting glass substrates by studying drops of polymer solutions of low viscosity and varying concentration $c$. To that end, we use solutions of polyethylene oxide (PEO, $M_W = 4.0 \times 10^6$ g/mol, Sigma-Aldrich) in distilled water. 
The solution concentration is varied from 0.01 to 2.0~wt\%, across $c^*\approx $~0.1--1~wt\%, the threshold over which polymer coils start to overlap (Supplementary Information~\cite{SI}). Each solution is mixed with a magnetic stirrer for at least 24 hours and degassed to get rid of air bubbles. 

Calibrations of the polymeric solutions used in this study are identical to those in \cite{Dekker2021}, repeated here for completeness in the Supplementary Information~\cite{SI}. 
The shear viscosities ($\eta$) of the solutions are measured using a rheometer (MCR 502 with CP50-1, Anton Paar), the surface tension ($\gamma$) and the relaxation times ($\lambda$) are deduced from pendant drop experiments, by fitting the drop shape and looking at the thinning of the connecting thread with the dispensing needle \cite{anna2001}, respectively. As reported in the SI, the polymer relaxation time $\lambda$ ranges from 1 to 50~ms in the explored range of concentrations $c= 0.01 - 2.0$~wt\%, and crosses the inertio-capillary time $\tau=\sqrt{\rho R^3/\gamma}\, \approx 3$ ms. Of particular importance is the values taken at the early stages of spreading of the ratio $\lambda/t$, a ``local instantaneous Deborah number", previously defined in \cite{Dekker2021} which reflects the deformation rate. This number takes values as high as 10$^2$ up to 10$^4$ at the smallest timescale resolved in our experiments when the polymer concentration is increased from 0.01 to 2.0~wt\%. A strong polymer stretching and elastic effects at the earliest stages of spreading are thus anticipated.  

As illustrated in figure~\ref{fig1}(a), the drop is slowly grown using a dispensing needle connected to a syringe pump (PHD 2000, Harvard Apparatus), pumping at a rate $<0.1\;\mu$L/min. The needle is placed about $2\rm\; mm$ above the horizontal substrate, in order to produce drops with radius $R\approx 0.9\rm\; mm$. Experiments are performed on clean transparent glass slides (supplied from Menzel-Gl\"aser), on which we measure static advancing contact angles of 52$^{\circ}\pm 4^{\circ}$.  It is well-known that  the exact nature of the substrate, specifically its wettability, has no influence on the early dynamics of a spreading droplet \cite{Winkels2012}.  Upon contact with the substrate, the liquid spreads. The deformations induced by the needle are local and far from the contact zone,  without disturbing the liquid bridge. The very slow increase in the drop volume enables approach velocities on the order of $1\;\mu$m/s, that is about three orders of magnitude smaller than typical bridge velocities The spreading dynamics is recorded both from the side and from the bottom using synchronized high speed cameras (Nova S12, Photron). The side-view camera is equipped with a zoom lens (Navitar lens), allowing for frame rates ranging from 100 to 200kfps and a spatial resolution of typically 2~$\mu$m/pixel. Typical snapshots of the early stage of the spreading of a drop with radius $R=0.9\pm 0.1$~mm are shown in Fig.~\ref{fig1}(b-c) for water and for a PEO solution of 2wt\%, respectively. Time increases from left to right with intervals of 125~$\mu$s between each image. Note that the upper half of the image represents the drop, while the lower half (below the white dotted line) is the reflection in the substrate.

Side views are complemented with synchronized bottom views. The bottom-view camera, mounted on an inverted microscope with a zoom lens (with 10$\times$ Olympus lens), shoots at 250kfps, which enables a better spatial and temporal resolutions than for the side-view. We display in Fig. \ref{fig1}(d) series of snapshots showing the liquid progression. Prior to contact, the thin air gap separating the drop from the solid produces an interference pattern, where fringes scroll as the drop approaches the substrate. After contact, the liquid quickly wets the solid: the dark area expands radially by approximately 200~$\mu$m in 50~$\mu$s, resulting in spreading velocities of about 5~m/s.

We focus on the spatio-temporal evolution of the bridge profile, $r(x,t)$, defined in Fig.~\ref{fig1} as the distance parallel to the substrate from the drop vertical axis of symmetry (located in $r=0$) and the liquid-air interface. We also denote as $r_0(t)$ the radius of the wetted area, which corresponds to $r_0(t)=r(0,t)$.
The profiles are extracted using a sub-pixel interface tracking code. Finding the precise moment of first contact between the drop and the substrate is crucial to determine the growth exponent $\alpha$. Here $t=0$ is determined by the bottom view, whose resolution is typically 5~$\mu$s, and further refined by extrapolating a power-law $r_0 \propto t^\alpha$, following the protocol as described in \cite{Dekker2021}.

\section{Rapid spreading dynamics}

\subsection{Temporal dynamics}

\begin{figure}[t]
	\centering
	\includegraphics[width=0.6\textwidth]{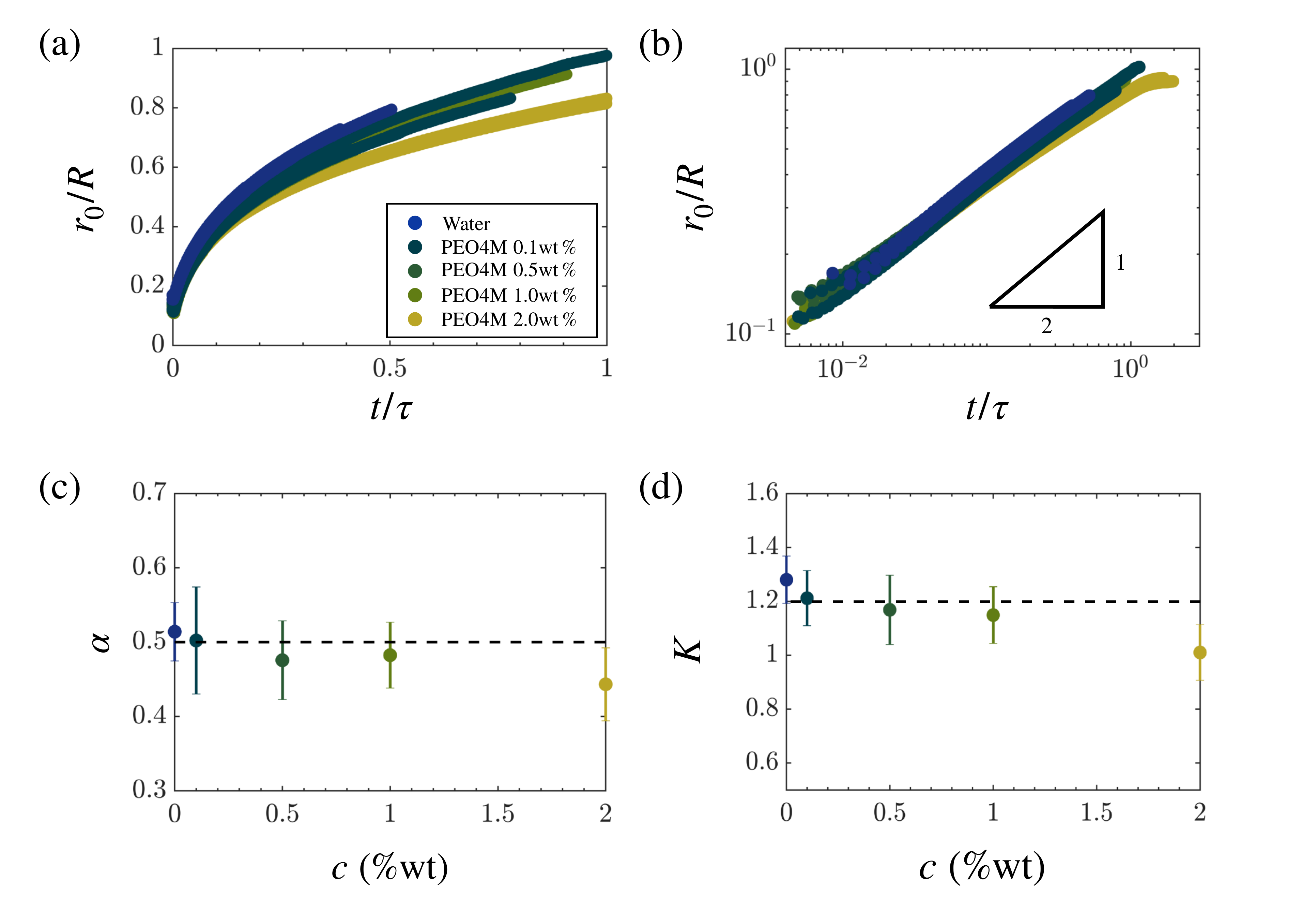}%
	\caption{Temporal spreading dynamics. (a) Contact radius $r_0$ (scaled by the drop radius $R$), as a function of time $t$ (scaled by the inertio-capillary time $\tau$) for water and PEO solutions of varying concentrations.  (b) Same data on a logarithmic scale. (c) Fitted exponent $\alpha$ extracted from $r_0/R= K(t/\tau)^{\alpha}$ as a function of $c$, the PEO concentration; the dashed line indicates the exponent expected for Newtonian liquids, $\alpha = 1/2$. (d) Prefactor $K$ as a function of a function of $c$ ; the dashed line indicates the value expected for Newtonian liquids, $K=1.2$ \cite{Biance2004,Winkels2012}. The error bars on $\alpha$ and $K$ arise from the averaging different experiments with the same liquid.}%
	\label{fig:time}
\end{figure}

The effect of polymers on the contact radius temporal growth is inconspicuous. Figure~\ref{fig:time} presents the temporal evolution of the contact radius $r_0(t)$ for spreading drops of varying polymer concentrations, both on linear (a) and logarithmic (b) scales. In view of the known behavior for water drops, the horizontal axis is scaled by the inertio-capillary time $\tau$ and the vertical axis by the drop radius $R$. Interestingly, the data exhibit a nearly perfect collapse for all polymer concentrations, especially at early times. Only at later times some differences become apparent. To further quantify this behavior, we fitted the  growth of the liquid bridge using a power-law of the type $r_0/R = K (t/\tau)^\alpha$. For water, the spreading exponent is known to be $\alpha=1/2$, while the numerical prefactor reported in literature is $K=1.2$ \cite{Biance2004,Winkels2012}. 
The fitted values of $\alpha$ and $K$ in our experiments on polymer solutions are presented in Fig.~\ref{fig:time}(c,d), for all tested concentrations. We find that the spreading exponent is consistent with $\alpha = 1/2$, for all polymer concentrations. Likewise, the  prefactor $K$ is in close agreement with the value measured for water. 

Remarkably, we find that the early stages of spreading of PEO solutions is nearly identical to that of pure water. Even though there is a very rapid motion, which certainly induces large polymer stretching, the temporal dynamics is hardly affected by the presence of polymer. At most, we observed a weak trend for the prefactor $K$, indicating a minor slowing down. We note that the highest concentration $c=2\mathrm{wt}\%$ departs slightly from the water exponent and prefactor, which can possibly be attributed to the fact that the concentration exceeds the overlapping concentration $c^*$ (estimated around $1\mathrm{wt}\%$ \cite{SI}); this mixture is clearly not in the dilute regime, which might be the cause why $\alpha$ differs slightly from the value observed at lower concentrations.  

An important comment to make is that the spreading of PEO solutions is very different from the spreading observed for Newtonian liquids of matching shear viscosity. For example, for $c=2 \mathrm{wt} \%$ the shear viscosity is $\eta\approx 10$~Pa.s (see SI), and the corresponding Newtonian drop would be deeply in the viscous regime. The  scaling in the viscous regime is linear with a logarithmic correction, as $r_0 \sim t \ln t$, which results into apparent spreading exponents close to 0.8 \cite{Eddi2013}. However, the data in Fig.~\ref{fig:time} shows no trend towards an increase of the exponent, from the inertial 1/2 toward the viscous scaling. We thus conclude that  in spite of the strongly enhanced shear viscosity of the polymer solutions, the behaviour for $\alpha$ and $K$ remains ``inertial" in nature. 

\subsection{Spatial structure}

\begin{figure*}[t]
	\centering
	\includegraphics[width=1\textwidth]{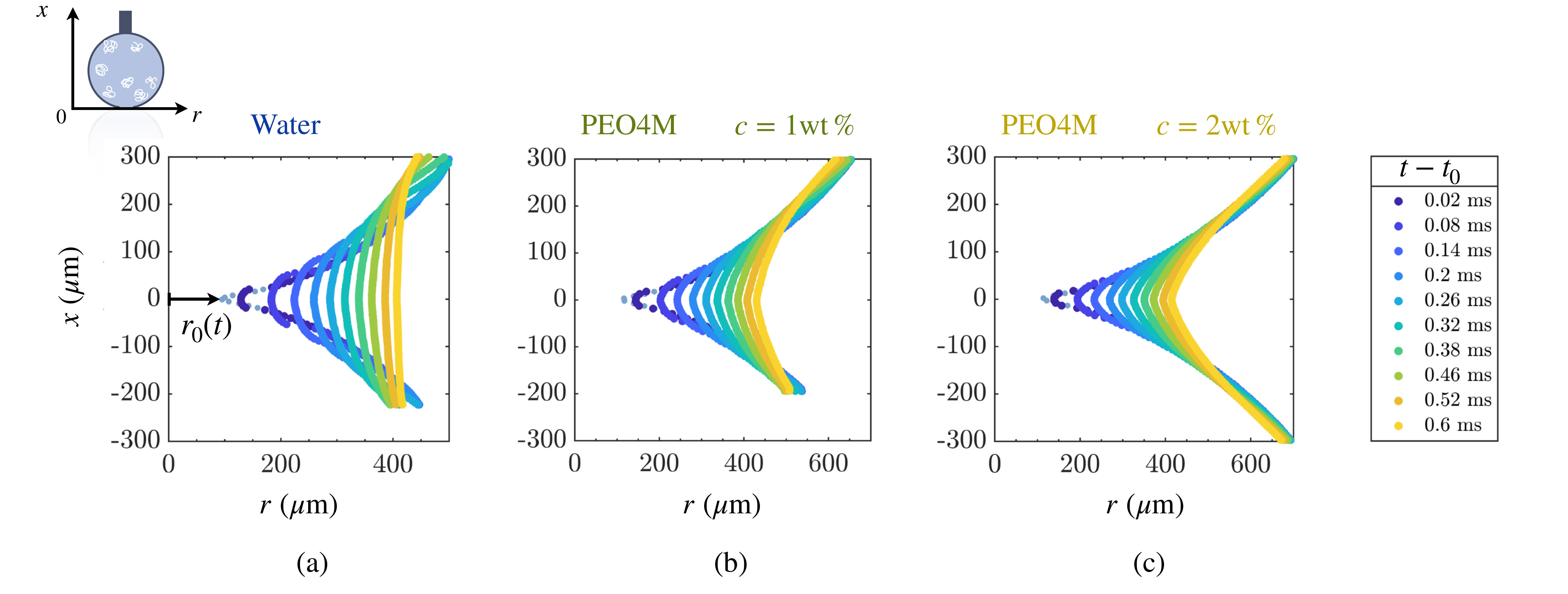}%
	\caption{Evolution of the bridge profiles (including the reflection in the substrate and the profile prior to contact in light blue) for spreading drops of (a) water, (b) PEO with 1wt\%, and (c) PEO with 2wt\%, shown at different times (with $t/\lambda =0.0007\to 0.02$ for the 1 wt\% PEO solution and $t/\lambda =0.0003\to 0.01$ for the 2 wt\%). Viscoelastic solutions exhibit a much sharper bridge than water.}
	\label{fig:profile}%
\end{figure*}

Though the \emph{temporal} dynamics of the spreading is hardly affected, the presence of polymer strongly modifies the \emph{spatial} structure of the bridge. This can be seen from the profiles $r(x,t)$ plotted in Fig.~\ref{fig:profile}, taken for different times after contact. The profiles appear nearly symmetric in the vertical direction, owing to the reflection in the substrate that is included in the plot. From these images it is evident that viscoelastic bridges are much sharper: while the water profiles are rather shallow (panel a), the interface exhibits a more pronounced curvature due to the addition of polymers (panels b,c). Remarkably, this effect is highly localised at the center of the bridge; the effect of polymer fades away at farther distance from the substrate. 

We now further quantify the sharpness of the bridge. First, we remark that the apparent contact angle of the interface with the substrate is very close to 90$^{\circ}$. This can be concluded from plotting $r(x,t)$ as a function of $x^2$. This plot gives a purely linear trend without any offset or asymmetry (data available in the SI \citep{SI}), 
which confirms that, on the scale of our experimental resolution, the apparent angle cannot be distinguished from 90$^{\circ}$. Subsequently, we can quantify the sharpness of the bridge by inspecting the bridge curvature $\kappa$ close to the substrate. The curvature is defined as $\kappa =  r_{xx}(t)$, which can be extracted at each instant from the slope of the curve of $r(x,t)$ versus $x^2$, in the vicinity of $x=0$ (cf. SI \citep{SI}). The curvature is of particular interest, since it gives direct information of the stress inside the liquid, according to Laplace's law of capillarity \cite{QuereBrochardPGG}. Hence, the enhanced curvature is a direct measure for the increase in stress due to polymer stretching.

\begin{figure}[tb!]
	\centering
	\includegraphics[width=1\textwidth]{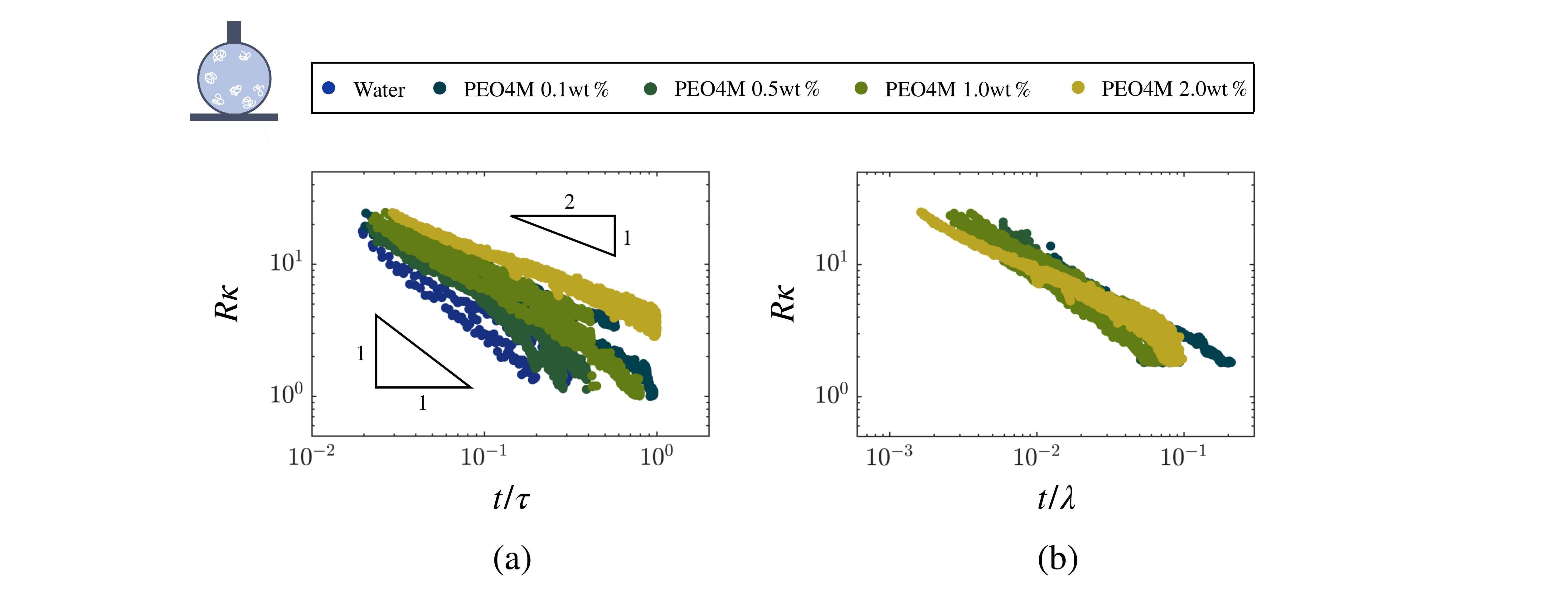}%
	\caption{Bridge curvature $\kappa$ (normalised by the drop radius $R$) plotted as a function of time $t$. (a) Time is normalised by the inertio-capillary time $\tau$. (b) Same data, but time is normalised by the polymer relaxation time $\lambda$. The suggested exponents are indicated as guide for eyes. }
	\label{fig:curvature}
\end{figure}

The resulting $\kappa(t)$ is plotted in Fig.~\ref{fig:curvature}, for different concentrations. In Fig.~\ref{fig:curvature}(a), the curvature $\kappa$ is made dimensionless with the drop radius $R$ \footnote{Normalisation with the dominant curvature, $R/r_0^2$, is perhaps more natural, but it markedly increases the noise in the data in the limit where $r_0\rightarrow 0$.}, while time is scaled with the inertio-capillary time $\tau$. For all concentrations, the bridge curvature is found to decay algebraically, with the smallest curvatures observed for water. The exponent does not appear to take on a distinct universal value, but changes with the nature of the liquid. We see a transition of the exponent from around $-1$ for water to lower apparent exponents, towards approximately $-1/2$ for the higher PEO concentrations.  Note that though rapid spreading of water drops is extensively studied, the bridge curvature is usually not considered; hence we are not aware that the behavior $\kappa \sim 1/t$ for water has been previously reported. In an attempt to account for the viscoelastic nature of the solutions, we also report the data by normalising time after contact with the polymer relaxation time $\lambda$, shown in Fig~\ref{fig:curvature}(b). This scaling gives a much better grouping of the data, which suggests that the dynamics of the curvature is indeed governed by polymer stretching, and its subsequent relaxation. The effective exponent is found to be  below $-1$, though we cannot identify a universal exponent.

\section{The spreading-coalescence analogy revisited}

We have thus found that the effect of viscoelasticity on rapid drop spreading is rather intricate. On the one hand, polymers hardly affect the motion of the contact line. Specifically, the spreading exponent remains identical to that for the initial spreading of water drops. On the other hand, the shape of the liquid-vapour interface in the bridge is strongly affected, as it exhibits much larger curvatures. 
This dual spatio-temporal influence of polymers is very strongly reminiscent to that observed recently in viscoelastic  coalescence \cite{Dekker2021}. In what follows we will explore the analogy between rapid spreading and coalescence  in the context of viscoelastic drops.

\begin{figure*}[b]
	\centering
	\includegraphics[width=1\textwidth]{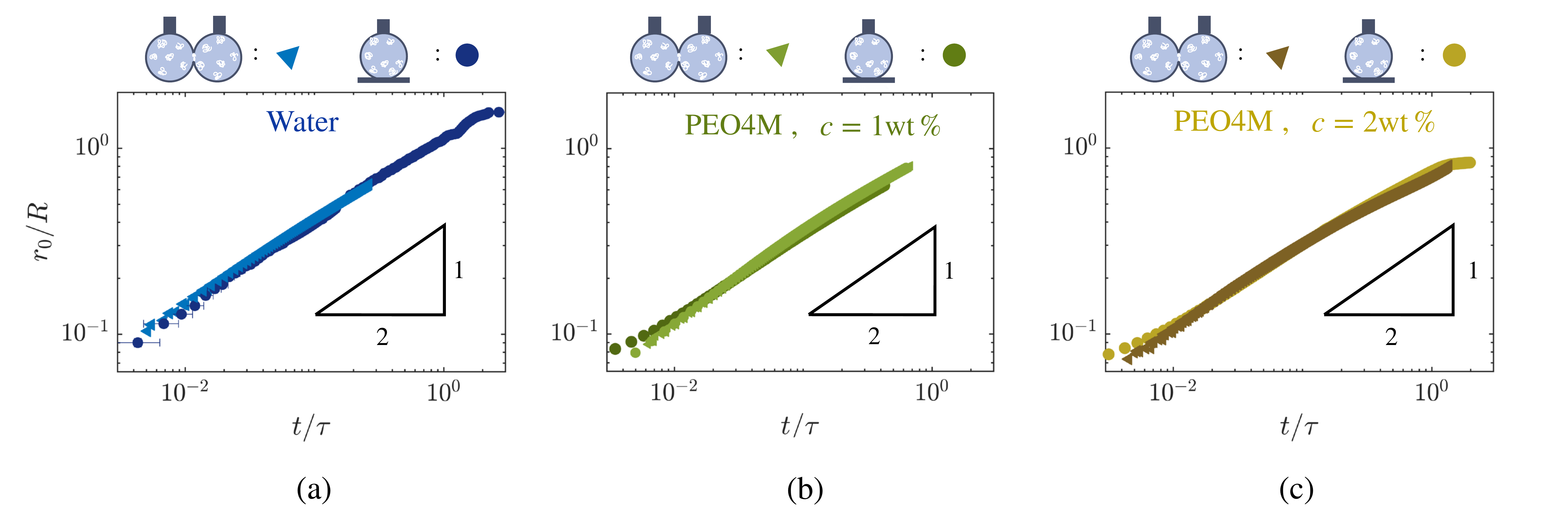}%
	\caption{Direct evidence for the spreading-coalescence analogy for the temporal evolution of the bridge. The three panels directly compare the dynamics of the bridge radius $r_0(t)$ for coalescence (triangles, taken from \cite{Dekker2021}) and spreading (circles). The data for spreading and coalescence are indistinguishable for (a) water, (b) PEO with 1wt\% and (c) PEO with 2wt\%.}
	\label{fig:SpreadingCoalescenceTime}%
\end{figure*}%

\subsection{Spatio-temporal dynamics of the liquid bridge}

We first focus on the time evolution of the bridge radius $r_0(t)$, reported in Fig.~\ref{fig:SpreadingCoalescenceTime}, for both spreading and coalescence. The data for coalescence are taken from the same experiments as those presented in~\cite{Dekker2021}. The results are again presented in dimensionless form, as $r_0/R$ versus $t/\tau$, for drops of water (panel a), PEO with 1wt\% (panel b), and PEO with 2wt\% (panel c). Clearly, in all cases, the data for coalescence (triangles) fall perfectly on top of the data for spreading (circles).  Hence, coalescence and spreading exhibit indistinguishible temporal dynamics, both in the Newtonian and viscoelastic cases. This implies that the presence of the substrate, in particular of the moving contact line, does not seem to have any influence on the wetting dynamics $r_0(t)$ during spreading; irrespectively of the presence of polymers inside the liquid.

\begin{figure*}[t]
	\centering\includegraphics[width=0.95\textwidth]{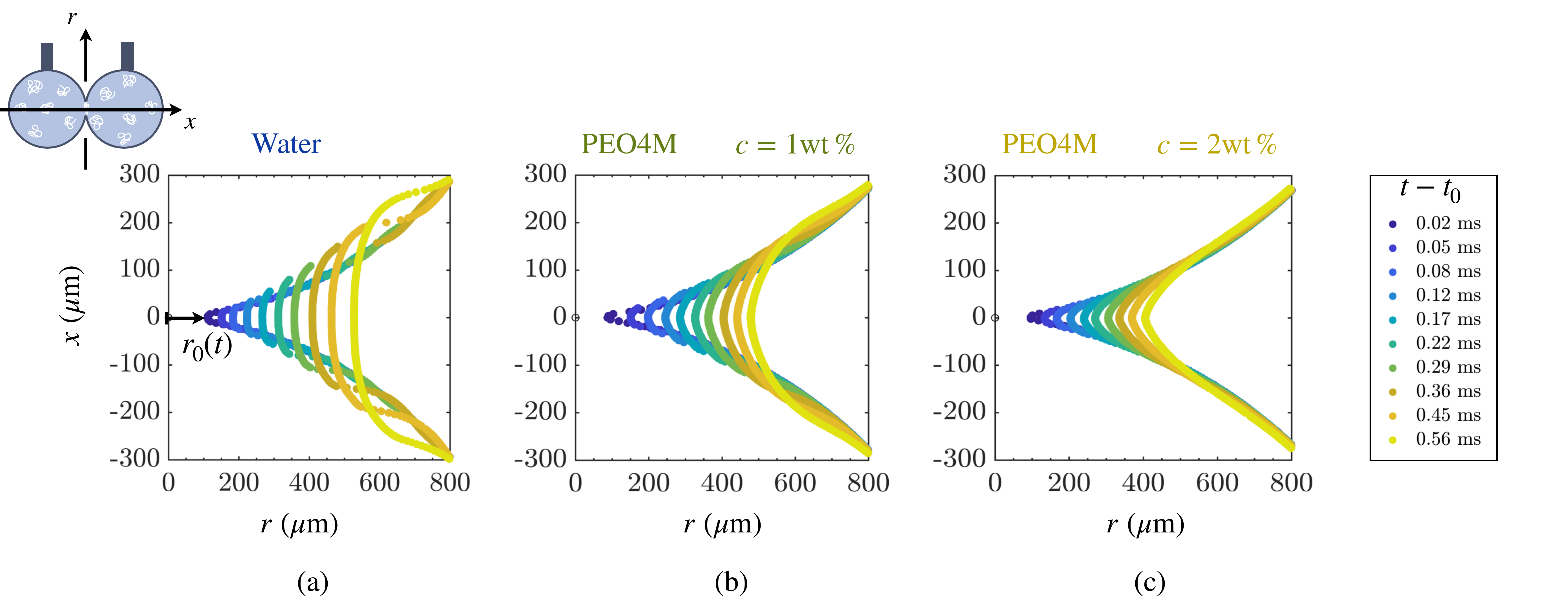}%
	\caption{Evolution of the bridge for coalescing drops of water, PEO with 1wt\% and PEO with 2wt\%. Bridge profiles (rotated by 90$^{\circ}$) are shown at different times matching those in figure~\ref{fig:profile}, with $t/\lambda =0.0007\to 0.02$ for 1 wt\% PEO and $t/\lambda =0.0003\to 0.01$ for 2 wt\% PEO. Viscoelastic solutions exhibit a much sharper bridge than water.}%
	\label{fig:SpreadingCoalescenceSpace}%
\end{figure*}%

On the face of it, the spatial structure of the liquid bridge observed during spreading is also analogous to that observed during coalescence of two drops. As shown in  Fig.~\ref{fig:SpreadingCoalescenceSpace}, the bridge profiles during coalescence are sharper when increasing the PEO concentration, in a way that is similar to drop spreading (for comparison see Fig.~\ref{fig:profile}). The two processes thus have in common that the stress inside the bridge is strongly enhanced due to the stretching of the polymer chains.  

\begin{figure}[t]%
	\centering\includegraphics[width=1\textwidth]{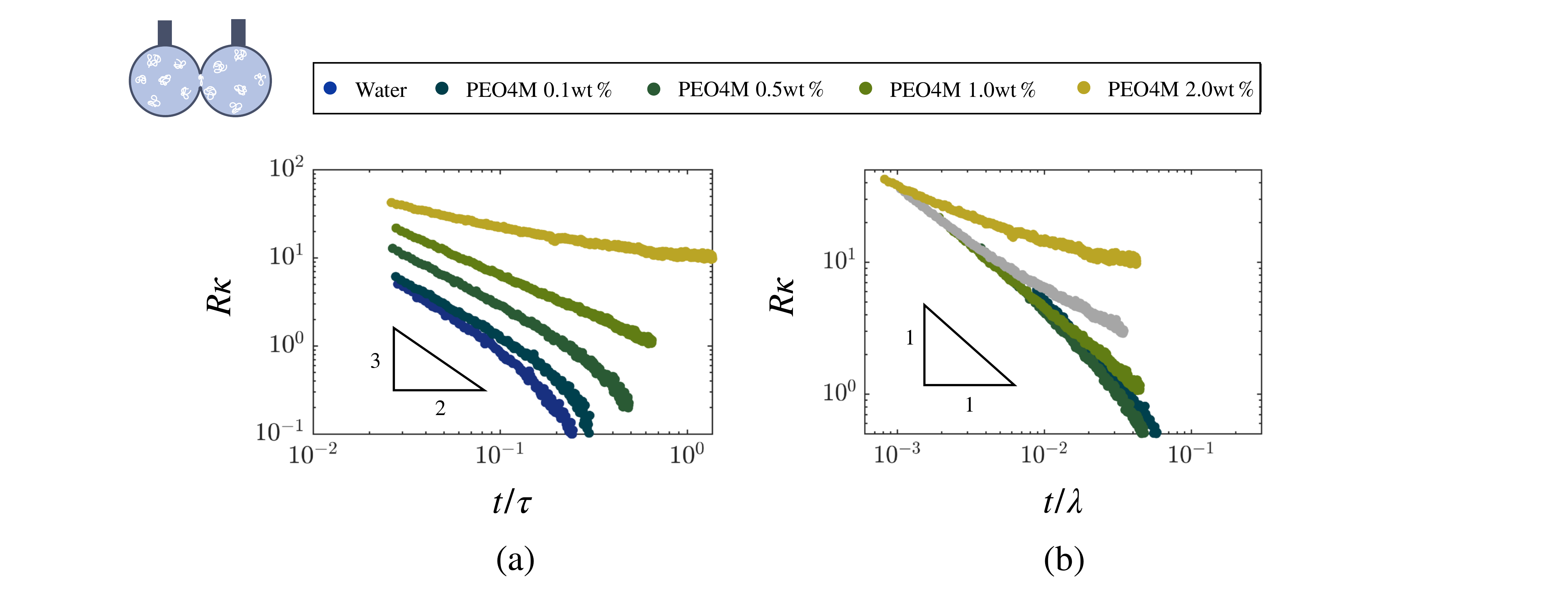}%
	\caption{Coalescence bridge curvature $\kappa$ (normalised by the drop radius $R$) plotted as a function of time $t$. (a) Time is normalised by the inertio-capillary time $\tau$. (b) Same data, but time is normalised by the polymer relaxation time $\lambda$. }
	\label{fig:curvatureCoalescence}%
\end{figure}

However, the central bridge curvature $\kappa(t)$ turns out to be essentially different. To make this apparent, we plot the temporal evolution of the curvature in Fig.~\ref{fig:curvatureCoalescence}, using the same rescalings as in Fig.~\ref{fig:curvature}. Let us first focus on the case of pure water. For coalescence, we find a scaling behavior that is close to $\kappa \sim 1/t^{3/2}$ [Fig.~\ref{fig:curvatureCoalescence}(a)]. However, for spreading of water drops we found $\kappa \sim 1/t$ [Fig.~\ref{fig:curvature}(a)]. The curvature data of coalescence clearly does not overlap witho that of spreading, which points to a breakdown of the spreading-coalescence analogy. A similar breakdown of the analogy is observed for polymer solutions. For this, one can directly compare Fig.~\ref{fig:curvatureCoalescence}(b) and Fig.~\ref{fig:curvature}(b), where time is normalised by the polymer relaxation time. For the case of coalescence, it appears that at early times there is a universal scaling $\kappa \sim 1/t$, which again differs substantially from $\kappa(t)$ observed during spreading. Hence, while there is a perfect spreading-coalescence analogy for $r_0(t)$, this analogy does not hold to the local curvatures. This implies that the local stresses at the interface are different for spreading as compared to coalescence.

\subsection{Is the bridge profile self-similar?}

\begin{figure*}[h!]
	\centering
	\includegraphics[width=0.95\textwidth]{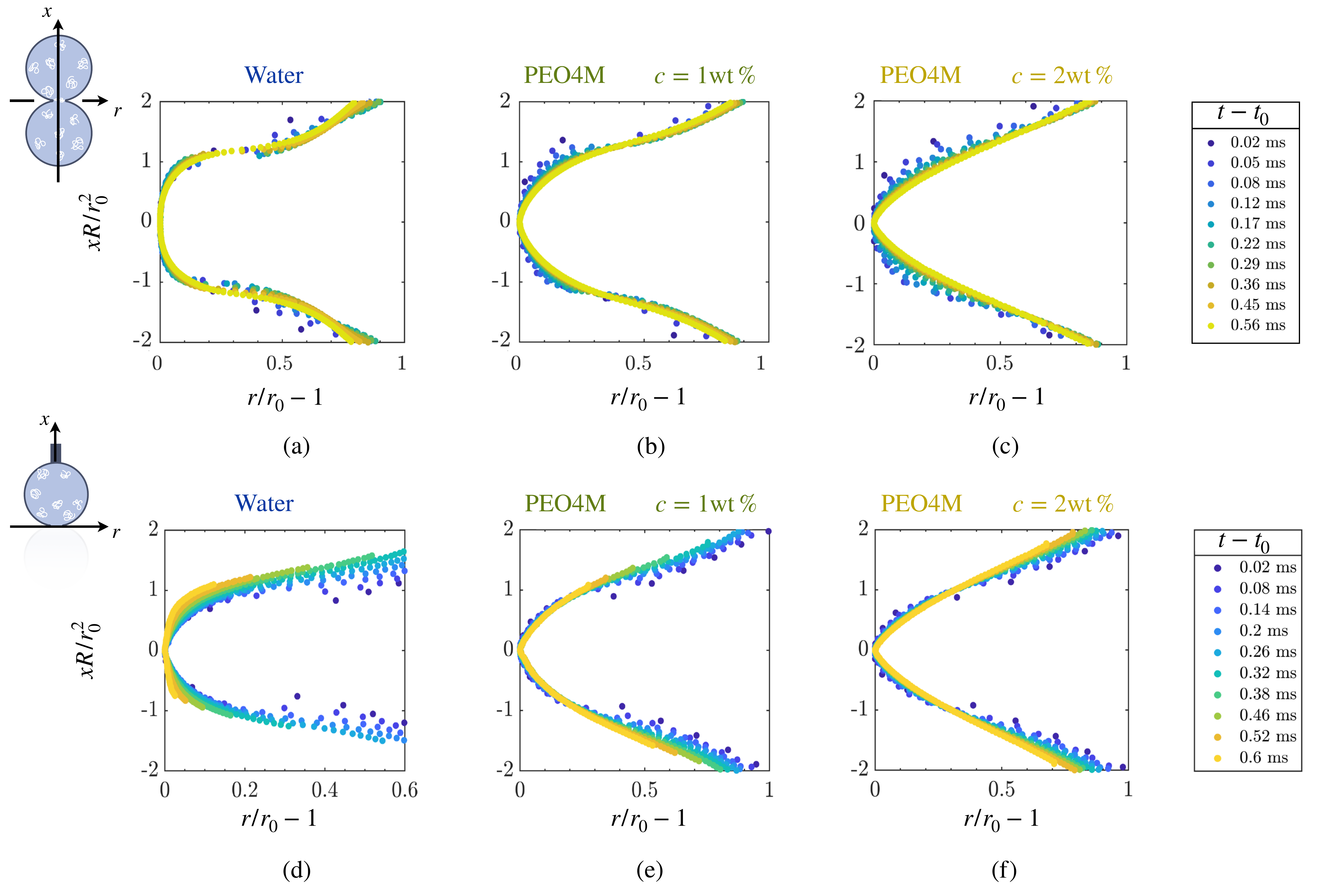}\label{fig:profileWaterScaling}
	\caption{(a-c) Bridge profiles obtained for coalescing drops of water, PEO with 1wt\% and PEO with 2wt\%, shown at different times, coinciding with Fig.~\ref{fig:SpreadingCoalescenceSpace}. Profiles are rescaled accordingly to the Newtonian self-similar scaling, that is by normalising the horizontal and vertical bridge extend $r$ by $r/r_0-1$ and $x$ by $xR/r_0^2$.  
(d-f) Bridge profiles obtained for spreading drops of water, PEO with 1wt\% and PEO with 2wt\%, shown at different times, coinciding with Fig.~\ref{fig:profile}, obtained by the same rescaling as in the top row.
}
\label{fig:SpreadingCoalescenceSpaceRescaling}%
\end{figure*}

To conclude the discussion on the spreading-coalescence analogy, we finally question whether the dynamics of the liquid bridge exhibits  self-similarity. For coalescence of Newtonian drops, it is known that the bridge profiles can be collapsed upon appropriate rescaling of spatial coordinates with time. Specifically, the scaling of the radial coordinate $r/r_0(t)$ and the axial coordinate $xR/r_0^2(t)$ gives a collapse of profiles during the coalescence of water drops \cite{Eddi2013}. In Fig.~\ref{fig:SpreadingCoalescenceSpaceRescaling} we therefore represent the spatio-temporal evolution of the bridge in viscoelastic spreading and coalescence according to this Newtonian scaling. 

We first focus on the case of pure water (left column of Fig.~\ref{fig:SpreadingCoalescenceSpaceRescaling}). Profiles for water coalescence nicely collapse (top-left panel), testifying that the dynamics is indeed self-similar. Specifically, the bridge profile follows a dynamics given by $r(x,t) = r_0 \mathcal H(xR/r_0^2)$, where $\mathcal H$ is a universal function. This data collapse for water coalescence also explains the observed scaling law for curvature. Namely, one evaluates $\kappa = r_{xx} \sim  R^2/r_0^3$, so that, using $r_0 \sim t^{1/2}$, we readily find $\kappa \sim 1/t^{3/2}$. Such a self-similarity is however not observed for the spreading of water drops (bottom-left panel). We recall that in this case we measured $\kappa \sim 1/t$, which explains why the attempted rescaling in Fig.~\ref{fig:SpreadingCoalescenceSpaceRescaling} naturally fails. This lack of self-similarity confirms that the spreading-coalescence analogy breaks down, even for water drops, when considering the spatial structure of the bridge. 

The central and right panels in Fig.~\ref{fig:SpreadingCoalescenceSpaceRescaling} correspond to bridge dynamics of viscoelastic drops.  For both coalescence (upper plots) and spreading (lower plots), we find that the rescaling gives a grouping of the profiles,  yet, with a collapse that is less convincing than for water coalescence (top-left).  
In particular the central region of the bridge does not collapse, which is in line with the differences in scaling of the bridge curvature $\kappa(t)$. A comparable breakdown with respect to the Newtonian self-similarity was found in \cite{Dekker2021} for the coalescence of \emph{sessile viscoelastic drops}, which are resting on a solid surface before merging. For these sessile drops, the polymer stress was proposed to scale as $\sigma \sim \bar \eta_\infty \dot r_0/r_0$, where $\bar \eta_\infty$ represents the extensional viscosity at high rates. For any power-law dynamics $r_0 \sim t^\alpha$, this implies an elastic stress $\sigma \sim 1/t$. As discussed in \cite{Dekker2021} for viscoelastic coalescence, the strong polymer stress in the bridge center gives rise to a new type of self-similarity, with scaling laws that arise from balancing polymer stress with surface tension. While for spherical viscoelastic drops the stress singularity during coalescence also appears to be $\sigma \sim 1/t$ (since $\kappa \sim 1/t$), we have not been able to identify a scaling that collapses the spherical coalescence (nor spreading). 

\section{Discussion} 

We have investigated the rapid spreading dynamics of a viscoelastic drop on a solid substrate, focusing on the spatiotemporal dynamics of the liquid bridge that forms immediately after contact. It was found that the time evolution of the contact radius was almost completely insensitive to the presence of polymer, and closely follows the purely inertial spreading law (\ref{eq:scalinglaw}). This, however, does not mean that the polymers do not affect the spreading: the interface profile of the liquid bridge is much sharper than the Newtonian case, owing to singular polymer stress. These observations are of interest for applications such as printing and spray deposition, where polymeric additives can be used to control the impact process.

Our experimental findings for rapid viscoelastic spreading are strongly reminiscent to those observed for viscoelastic drop coalescence. Indeed, just like for Newtonian drops, it is found that the bridge radius $r_0(t)$ is identical when comparing rapid spreading to coalescence. Importantly, however, we have found an important breakdown of the spreading-coalesence analogy. This is most directly appreciated from the bridge profiles in Fig.~\ref{fig:SpreadingCoalescenceSpaceRescaling}, by comparing the top row (coalescence) to the bottom row (spreading). From these rescaled profiles it can be seen the the bridges for spreading are systematically sharper than for coalescence -- pointing to a larger stress inside the bridge. In the case of pure water, we anticipate that this can be attributed to a viscous boundary layer associated to the substrate's no-slip boundary condition. In the case of polymer solutions, this effect could also be due to a viscous boundary layer or due to the enhanced stretching of the polymers by shear. While we have provided an interpretation for the observed curvatures during coalescence, the experimental findings for $\kappa(t)$ during spreading offer an interesting question for future studies. It also remains to be explained why, in all these cases, the temporal growth of the bridge is remarkably insensitive to the singular stress inside the central region of the bridge.

\vspace{0.25cm}
\begin{acknowledgments}
The authors thank Charu Datt and Walter Tewes for fruitful discussions. We acknowledge support from NWO Vici (No. 680-47-63), and from an Industrial Partnership Programme of NWO, co-financed by Oc{\'e}-Technologies B.V., University of Twente, and Eindhoven University of Technology. 
\end{acknowledgments}

\newpage
\bibliography{ViscoelasticSpreading}

\end{document}